\documentclass[10pt]{iopart}
\usepackage{tikz}
\usepackage{verbatim}
\usepackage{tikzscale}
\usetikzlibrary{arrows.meta,
 chains,
 positioning,shapes, decorations.pathreplacing,calligraphy,calc,
 shapes.geometric,arrows,decorations.pathmorphing,backgrounds,
 fit,positioning,shapes.symbols
 }
\usepackage{float}
\usepackage{fvextra}
\usepackage{adjustbox}
\usepackage{doi}
\usepackage{miller}
\usepackage{verbatim}
\usepackage{booktabs,threeparttable}
\usetikzlibrary{mindmap}
\usepackage{gensymb}

\begin{document}

\title[Batch learning boosted data-driven DFT]{Application of batch learning for boosting high-throughput \textit{ab initio} success rates and reducing computational effort required using data-driven processes}

\author{Robin Hilgers}
\address{Peter Gr\"unberg Institute and Institute for Advanced Simulation, Forschungszentrum J\"ulich and JARA, 52425 Jülich, Germany\\
Department of Physics, RWTH Aachen University, Aachen, Germany
}\ead{robin.hilgers@rwth-aachen.de}
\author{Daniel Wortmann}
\address{Peter Gr\"unberg Institute and Institute for Advanced Simulation, Forschungszentrum J\"ulich and JARA, 52425 Jülich, Germany}
\author{Stefan Bl\"ugel}
\address{Peter Gr\"unberg Institute and Institute for Advanced Simulation, Forschungszentrum J\"ulich and JARA, 52425 Jülich, Germany\\
Department of Physics, RWTH Aachen University, Aachen, Germany
}

\begin{indented}
\item[]\today
\end{indented}
\begin{abstract}
The increased availability of computing time, in recent years, allows for systematic high-throughput studies of material classes with the purpose of both screening for materials with remarkable properties and understanding how structural configuration and material composition affect macroscopic attributes manifestation. However, when conducting systematic high-throughput studies, the individual \textit{ab initio} calculations' success depends on the quality of the chosen input quantities. On a large scale, improving input parameters by trial and error is neither efficient nor systematic. We present a systematic, high-throughput compatible, and machine learning-based approach to improve the input parameters optimized during a DFT computation or workflow. This approach of integrating machine learning into a typical high-throughput workflow demonstrates the advantages and necessary considerations for a systematic study of magnetic multilayers of 3$d$ transition metal layers on FCC noble metal substrates. For 6660 film systems, we were able to improve the overall success rate of our high-throughput FLAPW-based structural relaxations from $64.8 \%$ to $94.3\ \%$ while at the same time requiring $17\ \%$ less computational time for each successful relaxation. 
\end{abstract}
\vspace{2pc}

\ioptwocol
\maketitle
\section{Introduction}

Ultrathin magnetic multilayer film systems represent a tunable platform~\cite{surf2,tun} for spintronics applications, as they exhibit enhanced magnetic properties due to the more pronounced presence of surface effects~\cite{surf,surf2}. This leads to \textit{e.g.} the magnetic moments to be increased in a 2-dimensional film when compared to a 3-dimensional bulk of similar composition~\cite{Bluegel2D}. Motivated by the emerging magnetic phenomena, including such as room-temperature stable Skyrmions~\cite{tun}, giant magnetoresistance~\cite{GMR}, and the anomalous hall effect~\cite{AHE}, systematic high-throughput studies examining the film's magnetic properties in relation to \textit{e.g.} the corresponding film's composition, layer ordering, and layer thickness. In our case, we restricted this study to film systems with - at most - 3 layers of 3 $d$ transition metals on five layers of the FCC noble metals as substrates. This opens a phase space of 6660 structures, initialized as magnetic films and relaxed. This initialization is necessary as not all 3 $d$ transition metals are magnetic in bulk systems but might become magnetic in ultrathin film systems due to the mentioned surface effects. 

Systematic studies are essential when it comes to materials screening in search of a specific combination of material properties, but also when it comes to understanding the tuneable parameters that impact desirable material features, which can be used to effectively design a compound based on that knowledge for the material to exhibit very distinct magnetic effects or configurations. However, within high-throughput studies, typically, many \textit{ab initio} calculations are required to determine an individual property (such as \textit{e.g.} half-metallicity~\cite{halfMet}, relaxed structure~\cite{str}, critical temperature~\cite{tcFilms}, etc.). However, converging a single self-consistent Density Functional Theory (DFT) calculation of a magnetic system can be challenging. Relaxing the computed structures adds another level of complexity. Now, wrapping both problems into a high-throughput context again provides its own distinct problems. High-throughput specific problems include \textit{e.g.} choosing appropriate starting parameters (\textit{e.g.} initial magnetic moments, starting inter-atomic distances, etc.) for the computed – and potentially very diverse – structure configurations and the necessity for a systematic – and in the best case automated – approach to tweaking failed workflows/calculations while maintaining consistency of the results. 

However, finding solutions for the mentioned problems is crucial for understanding subclasses of materials like, in our example, layered thin-film systems because in publications, often there are only a few systems per subclass examined rather than a systematic search being performed on the respective subclass. This leads to much knowledge centered around a few materials or compounds, while the bigger picture can remain unclear.
 To learn about the bigger picture of materials subclasses, high-throughput frameworks and workflows represent a well-suited method. 

In the following, we showcase a machine learning (ML) based method that we developed to boost convergence rates of high-throughput workflows/calculations, reduce the required iterations, and hence reduce the overall energy consumption that emerges from the related use of HPC systems. We refer to the method as DFT integrated ML (DFT IntML). This approach is suitable for high-throughput workflows/setups where a quantity for which you have an initial guess at the beginning of the workflow is optimized during the execution of the workflow. This approach fits in the category of data-driven materials design methods and resembles an application case of the batch learning method. We applied this methodology to symmetrical 2-dimensional films of magnetic 3$d$ transition metal layers on FCC noble metal substrates. The database which resulted from the FLAPW~\cite{FLAPW,FLAPWCit}
calculations, which have been performed using the \texttt{\uppercase{Fleur}} code~\cite{fleurWeb,fleurCode} within the high-throughput compatible Automated Interactive Infrastructure and Database for Computational Science (AiiDA) framework~\cite{AiidaA,AiidaB} together with the AiiDA-FLEUR plugin~\cite{aiidaFleur,Masci}, is publicly available~\cite{2DDatapub}.
Additionally, the code used to analyze and visualize the data, train and evaluate the ML models, and apply the DFT IntML input optimization approach is published on Zenodo~\cite{2DCodepub}.

The methodology, as well as our specific application case, is presented in-depth in the following.

\section{Methods \& Materials}
\subsection{Film Construction}
\begin{figure*}[t]
  \includegraphics[width=\linewidth]{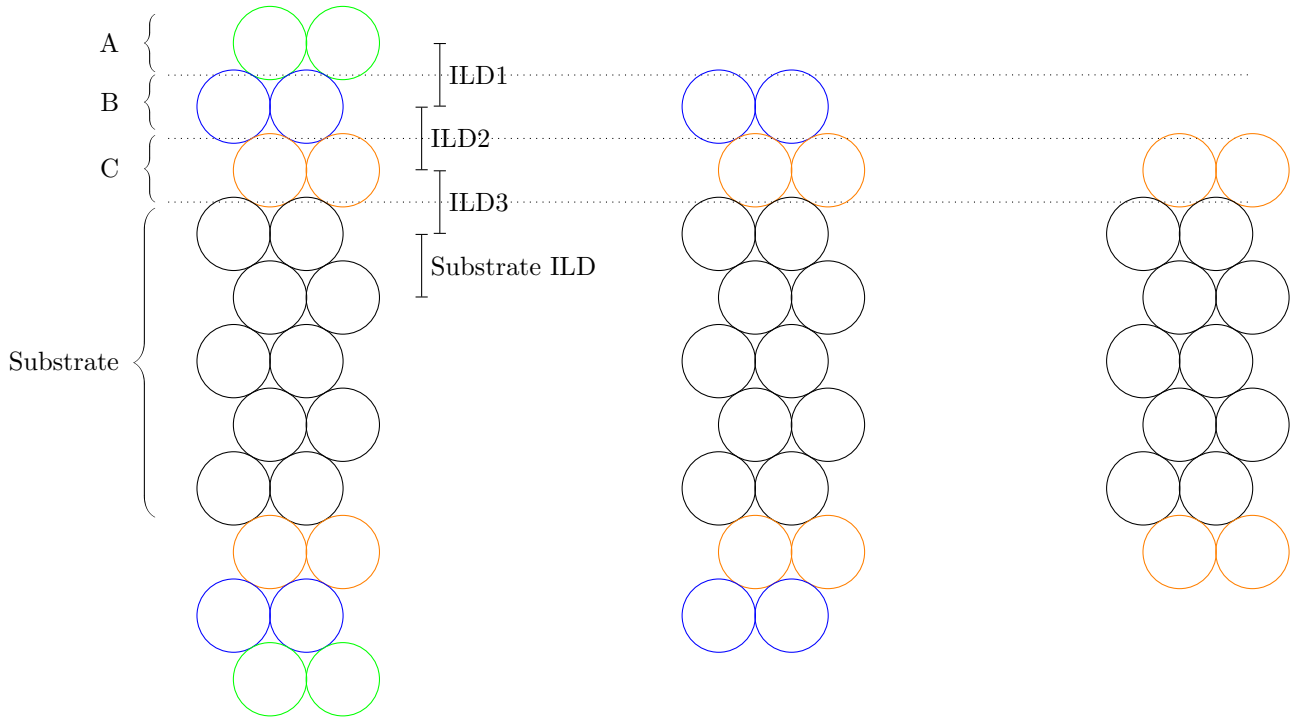}
  \caption{Depiction of the structural setup (lateral view) for films with 3, 2 and a single (colored) magnetic layer on the (black) FCC (100) noble-metal substrates, atomic sites naming conventions and graphical depiction of the ILDs order convention. }\label{2DSetup}
\end{figure*}

Constructing several layered \hkl[001] FCC film systems on a high-throughput scale for a systematic search is a demanding task on its own. We were able to use existing AiiDA-FLEUR~\cite{aiidaFleur} 
workflows developed at our institute,
 which construct the films and start the relaxation process using the AiiDA framework~\cite{AiidaA,AiidaB,Vasily}.
In Fig.~\ref{2DSetup}, the structural setup and naming conventions for atomic sites and interlayer distances (ILDs) of our films are displayed. The layers A, B, and C, representing the magnetic layers of 3$d$ transition metal elements, are stacked on top of the five substrate layers on each side of the substrate to create a symmetric film. Of course, setting up the substrate layers in the corresponding substrate layer system and subsequently adding magnetic layers on the top and the bottom of these substrate layers dictates the lattice system. The magnetic layers are placed in the substrate lattice system with the substrate in-plane lattice constant. A symmetrical film was chosen since this increases the number of symmetry operations applicable to the resulting structure and subsequently improves computational efficiency. A symmetric film would also have been achievable using an even number of substrate layers; however, using an even number of substrate layers would conclude that either inversion symmetry or the z-reflection would be lost. The loss of inversion symmetry would introduce complex numbers within the DFT computation algorithm, which is unfavorable. Hence, to maintain z-reflection and inversion symmetry, we conducted this study using an odd number of substrate layers within the symmetric films.

The mentioned workflow contains an option to estimate ILDs by using the mean bond length of both neighbors of all compounds contained in the Materials Project~\cite{MatProj} database. Using this option to set up the initial structure, the initial ILD guess for neighboring layers is based on the corresponding atom pair's average bond length of the bonds in the Materials Project~\cite{MatProj} database. This, however, means that no additional information about in-plane neighbors or next-nearest-neighbors is used in the first bond length guess computation. Since we are using the ILDs in a film setup, we are scaling the ILD between the atoms A and B on the very outside of the film by multiplying our workflow guess by a factor of 0.95, as it is known that outer layer boundaries tend to compress to some degree.\label{MatProjGuess}
\subsubsection{Substrate \& Layer Selection}
As substrate elements, we decided on Pt, Au, Ag, Ir, Pd, and Rh, as noble metals in films are known to have good adhesive capabilities to add metal layers to them.
 Currently, our film structure setup workflow~\cite{aiidaFleur} supports the setup of FCC and BCC substrate lattices.

As we are particularly interested in magnetic multilayers, we decided to go with a class of elements likely to exhibit magnetic properties in a film setup: the 3$d$ transition metals~\cite{Bluegel2D}.
 The 3$d$ transition metal elements are Sc, Ti, V, Cr, Mn, Fe, Co, Ni, Cu, and Zn. Allowing site A to be unoccupied and sites A and B to be commonly unoccupied, combined with the number of 6 substrate elements, enables the construction of 6660 film systems. 

The choice for a substrate thickness of 5 substrate layers was made since we previously observed that relaxed ILDs for magnetic layers are already converged for a substrate layer count of 3, down to a change below $5\ \%$ compared to increasing layer thicknesses. Hence, we decided to go with five substrate layers as a trade-off between accuracy and computational requirements to ensure the results are converged, and the resulting ILDs and structures would be close to those of larger substrate thicknesses.\label{Const}
\subsubsection{Relaxation Workflow}
\begin{figure*}[ht]
\resizebox{\linewidth}{!}{    
\begin{tikzpicture}[
   node distance = 5mm and 7mm,
      start chain = going right,
 round/.style = {draw,
    rounded corners=0.5cm,
    minimum width=2.5cm,
    minimum height=1cm,
    text width=30mm, align=center, font=\linespread{0.8}\selectfont},
        disc/.style = {shape=cylinder, draw, shape aspect=0.3,
                shape border rotate=90,
                text width=16mm, align=center, font=\linespread{0.8}\selectfont},
  rect/.style = {draw,
    rounded rectangle,
    rounded rectangle arc length=45,
    minimum width=2.5cm,
    minimum height=1cm,text width=30mm, align=center, font=\linespread{0.8}\selectfont},
  alg/.style = {draw, align=center, font=\linespread{0.8}\selectfont},     arrow/.style = {thick, -Stealth},
  decision/.style = {diamond, aspect=1.5, draw, fill=green!30,
                    minimum width=3cm, minimum height=1cm, align=center,
                    on chain, join=by arrow},
                    ]
\node (n1) [rect] {CreateFilmWC};
\node (n10) [round,right = 10mm of n1] {EOS};
\node (n0) [round,right = 10mm of n10] {Relaxation };
\node (n2) [round,   right = 10mm of n0]  {SCF Calculation};
\node (n5) [decision,   right = 10mm of n2]  {Converged \\ SCF?};
\node (n3) [decision,   right = 10mm of n5]  {Converged \\ Relax?};
\node (n6) [disc,   right = 10mm of n3]  {Structure and\\ Magnetism};
\draw[arrow] (n1.east)  -- (n10.west);
\draw[arrow] (n10.east)  -- (n0.west);
\draw[arrow] (n0.east)  --   (n2.west);
\draw[arrow] (n2.east)  -- (n5.west);
\draw[arrow] (n3.east)  --  node[pos=.25,above]{Yes}  (n6.west);
\draw[arrow] (n5.east)  --  node[pos=.25,above]{Yes}  (n3.west);

\draw[arrow] (n5.north) |- ++ (0,0.00003) -|  node[pos=.25,above]{No} (n2.north);
\draw[arrow] (n3.north) |- ++ (0,1) -| node[pos=.25,above]{No} (n0.north);
    \end{tikzpicture}\unskip}
\caption{Simplified flowchart of a film relaxation workflow in the AiiDA-FLEUR implementation.}\label{RelaxWF}
\end{figure*}
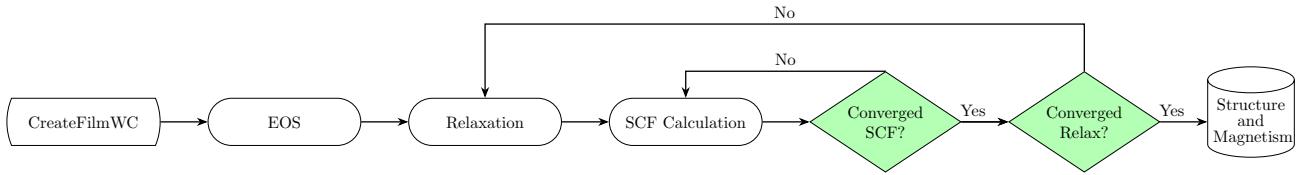

The whole relaxation workflow is depicted simplified in Fig.~\ref{RelaxWF}. 
 To choose appropriate substrate lattice constants - which are not subject to relaxation during the workflow - an EOS computation has been performed on the substrate lattices. An initial substrate lattice constant was determined from the Materials Project~\cite{MatProj} bond length guesses, which was used to set up multiple EOS calculations with scaled lattice constants and determine the substrate lattice constant with the lowest total energy. Table~\ref{LatCon} depicts the determined substrate lattice constants, which are in excellent agreement with GGA FLAPW reference values~\cite{DeltaProject}.  

\begin{table*}[h!t]
 \centering
 \begin{tabular}{lllllll}
\toprule
Element& Rh & Pd & Ag & Ir & Pt &Au\\
\midrule
Lattice Constant in \AA& 3.83& 3.94  &4.14  &3.87  & 3.97 & 4.15  \\
\bottomrule
\end{tabular}
 \caption{Substrate FCC lattice constants acquired by finding the minimum energy value computed within the EOS calculations using the scaled initial lattice constant guess.}
 \label{LatCon}
\end{table*}
\vspace{1cm}

After the EOS evaluation and substrate lattice constant determination, the film is constructed first only consisting of substrate atoms using the EOS resulting lattice constant. The final film setup step is replacing the layers chosen to be occupied by a magnetic atom layer. This includes re-scaling the ILDs so that the initial guess for the bond length between the neighboring atoms is matched. 

After the film setup, the relaxation loop itself is started. Each relaxation step requires a self-consistent field (SCF) calculation to be converged before the structure is adjusted according to the forces resulting from the SCF calculation. In this study, we allowed a total of (at most) 100 relaxation steps with 100 SCF iterations each. A single SCF calculation was allowed to restart nine times if it failed for a reason that another run could fix (SCF convergence, process externally killed, etc.). We enforced a convergence criterion of $10^{-3}\frac{\mathrm{m e^-}}{a_0^3}$ for the charge density distance of the SCF calculations and $5\cdot 10^{-5}\frac{\mathrm{Ha}}{a_0}$ maximum force.
The latter can be considered a rigorous criterion, as we wanted to prevent “falsely” relaxed structures. As the substrate lattice has already been optimized in the EOS calculation, the substrate is kept fixed, as well as the magnetic layers $x$ and $y$ position coordinates. The relaxation is performed along the $z$-axis. This means the previously mentioned ILDs are the only relevant positional parameters that change.  

Each successful relaxation workchain results in a relaxed structure. Alongside the structure, the resulting magnetic configuration, together with additional (Total energy, etc.) and metadata (Number of relaxation steps, total number of SCF iterations, etc.), is stored within the AiiDA database.\label{strucini} 
\subsubsection{Initial Magnetic Setup}
Some elements in our selection of magnetic layers are known to tend to be non-magnetic (\textit{e.g.} Cu). However, as we are particularly interested, each atom in the magnetic layers is assigned an initial spin-polarization of $1\ \mathrm{\mu B}$. This is referred to as the magnetic initial guess. It is essential to avoid the construction of entirely non-magnetic films from the beginning, as the \texttt{\uppercase{Fleur}} code would maintain this symmetry by not spin-polarizing the system. 

A consequence of this choice is that all films – regardless of their composition and structure – are initialized as ferromagnetic. \label{magini}

\subsection{DFT Integrated ML}
DFT IntML is a form of batch learning combined with the \textit{ab initio} simulations approach. Here, batch learning means an ML model is trained on a database that contains a subset of all possible data entries, and the model can then predict the remaining data entries based on the learned subset. Still, once more data for training is available, it is used to retrain the model, which results in updated predictions. For our case, this means: We train a model to predict target quantities\footnote{Note: In this paper, we apply this to the initial structural and magnetic configuration. But in principle, this can be applied to any quantity which is both input as well as output of an \textit{ab initio} calculation.} from \textit{ab initio} calculations outputs using descriptors that do not require a DFT calculation beforehand (\textit{e.g.} atomic numbers) – knowing there are more DFT calculations that we want to perform – then we can make predictions with the trained model and acquire estimates for this unseen data. These predictions are then used as an improved starting point for the \textit{ab initio} setup, potentially reducing the computational time required and elevating the chances of success for this particular calculation. 
As with each DFT IntML iteration, additional data was available for model training and has been used to predict the input parameters for the remaining set of film structures to be relaxed. This iterative approach represents a form of batch learning. 

There are already molecular dynamics simulations carried out entirely based on ML models. However, we use ML models within the DFT IntML scheme without losing the theoretical backing provided by DFT, as we used the ML predictions as inputs in the subsequent following \textit{ab initio} workflows. Hence, DFT IntML is not replacing \textit{ab initio} methods but complementing them for increased success rates and efficiency. 

\begin{figure}[ht]
\resizebox{0.5\textwidth}{!}{    
\begin{tikzpicture}[tips=proper,
   node distance = 5mm and 7mm,
      start chain = going right,
 round/.style = {draw,
    rounded corners=0.5cm,
    minimum width=2.5cm,
    minimum height=1cm,
    text width=30mm, align=center, font=\linespread{0.8}\selectfont},
    roundsmall/.style = {draw,
    rounded corners=0.5cm,
    minimum width=1cm,
    minimum height=1cm,
    text width=15mm, align=center, font=\linespread{0.8}\selectfont},
    disc/.style = {shape=cylinder, draw, shape aspect=0.3,
                shape border rotate=90,
                text width=20mm, align=center, font=\linespread{0.8}\selectfont},
  rect/.style = {draw,
    rounded rectangle,
    rounded rectangle arc length=45,
    minimum width=2.5cm,
    minimum height=1cm,text width=30mm, align=center, font=\linespread{0.8}\selectfont},
  alg/.style = {draw, align=center, font=\linespread{0.8}\selectfont},     arrow/.style = {thick, -Stealth},
  line/.style={thick},
  decision/.style = {diamond, aspect=1.5, draw, fill=green!30,
                    minimum width=3cm, minimum height=1cm, align=center,
                    on chain},
                    ]
\node (n0) [alg] {Trained \\ Production Model};          
\node (n1) [disc,right = 10mm of n0] {Predicted Input Quantities};
\node (n2) [rect,right = 10mm of n1] {High-Throughput Workflow};
\node (n3) [disc,right = 16.5mm of n2] {Output Data Point};
\node (a0) [disc,above = 5mm of n0] {Not computed structures};
\node (a1) [inner sep=0,minimum size=0,right  = 5mm of n0] {};
\node (n4) [disc,below = 10mm of n3] {Results Database};
\node (n5) [decision,left = 10mm of n4] {Significant \\accumulated amount of\\ additional data?};
\node (n6) [disc,below = 10mm of n5] {Train/Test Split};
\node (n7) [alg,left = 2mm of n6] {Train Model};
\node (n8) [decision,left = 7mm of n7] {Test performance \\better than\\ production Model?};
\node (n9) [round,above = 12mm of n8] {Retrain production model on complete data set};
\node (n10) [roundsmall,right = 5mm  of n9] {Get more data};
\draw[line] (a0.east) -| (a1.north);
\draw[arrow] (n0.east)  -- (n1.west);
\draw[arrow] (n1.east)  -- (n2.west);
\draw[arrow] (n2.east)  -- (n3.west);
\draw[arrow] (n3.south)  -- (n4.north);
\draw[arrow] (n4.west)  -- (n5.east);
\draw[arrow] (n5.south)  --  node[pos=.25,right]{Yes}  (n6.north);
\draw[arrow] (n6.west)  -- (n7.east);
\draw[arrow] (n7.west)  -- (n8.east);
\draw[arrow] (n6.south)  |- (n8.south);
\draw[arrow] (n8.north)  --  node[pos=.25,right]{yes}  (n9.south);
\draw[arrow] (n9.north)  -- (n0.south);
\draw[arrow] (n8.north east)  -| node[pos=.25,above]{No}  (n10.south);
\draw[arrow] (n5.south)  -| node[pos=.25,above]{No}  (n10.south);
\draw[line] (n10.west)  -| (a1.south);
    \end{tikzpicture}\unskip}
\caption{Depiction of the DFT IntML data and workflow. Cylinders denote different stages in which the processed data is located. Green diamonds symbolize decisions in the workflow.}\label{IntML}
\end{figure}
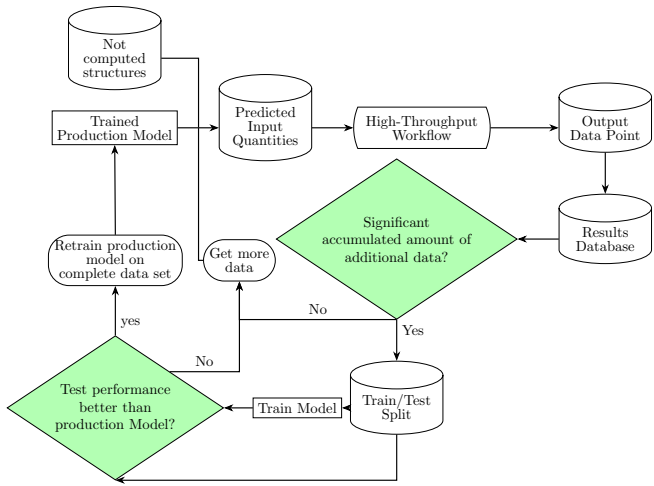

The entire process and data flow of this method is shown in Fig.~\ref{IntML}.
The question of when enough data has accumulated to process and evaluate a new model does not have a simple answer. In a continuous data stream setting, frequently reevaluating the production model and retraining, including the additional data, could be desirable. This would also be desirable, as this would maximize predictive accuracy due to the quick usage of the acquired data; if the goal is to save as much HPC computation time as possible, most ML methods (excluding artificial neural networks) are not costly in the training and evaluation phase. If the main goal is to maximize the success/convergence rate, it is enough to perform learning steps as larger batches of data become available. This was also the case for our application; We predicted the remaining data set and computed every missing structure for each consecutive batch. 

Our data-driven approach to high-throughput calculations requires a database to build on. As described previously, we used initial guesses for the ILDs and magnetic moments of the outer non-substrate layers to perform an initial set of calculations, which outputs we used for model training and prediction to replace first the magnetic moment guess and in a subsequent step also the ILD guesses.
 However, one also has to consider that the initial guess can affect the outcome of the DFT calculations, as for a single film structure, multiple magnetic configurations could represent (meta) stable states of the systems. Hence, the choice of magnetic initialization could affect the resulting magnetic configuration, which then could transfer to the ML model if trained on this data. This seems not to be the case for the chosen ferromagnetic initialization, as we observed non-magnetic, ferrimagnetic, and anti-ferromagnetic configurations that emerged from the batch using only the initial ferromagnetic moment guesses. 

\subsection{Model Choice for DFT IntML}
There is no general rule for choosing an ideal model. The famous “no free lunch” theorem~\cite{Lunch} also includes that it is unknown which model will work best before the models have been trained and evaluated. 

\subsection{Starting Point for DFT IntML}
Generally speaking, the ideal moment to start the use of DFT IntML is as soon as other methods to determine the target quantity are outperformed in terms of the average absolute deviation to the successfully converged \textit{ab initio} result, as this implies that the ML-based prediction provides an improved starting point, compared to the previously used method.
 Depending on the data complexity, the chosen ML model, and the number of features, the amount of data required to outperform other - potentially data agnostic - methods can drastically vary. 
In our case, the guesses of the initial ILDs and magnetic moments are, as described previously, independent of the amount of data acquired. Hence, the mean absolute error (MAE) of the DFT-based workflow-determined values and the guessed values – on average – is constant concerning the total number of available data entries. Therefore, as soon as the MAE of DFT IntML predicted input quantities are lower than this constant error, the high-throughput process could have been continued using the DFT IntML predictions. However, we computed the very first batch\footnote{Which we labeled batch 0. This batch contains over 4000 entries, in our case.} entirely with the initial guesses to determine the success-boosting effect of DFT IntML in comparison to the use of the previously described initial guesses.

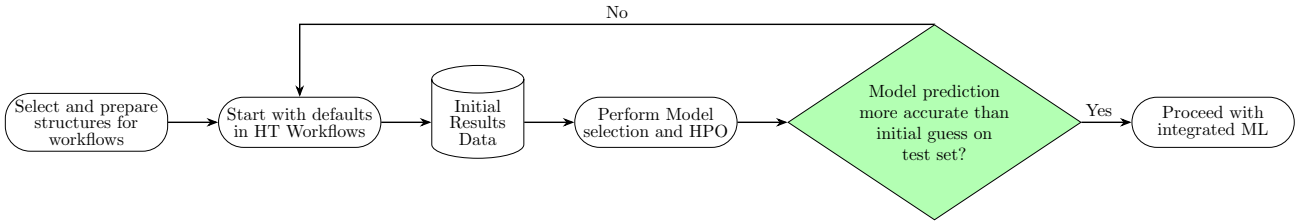
\begin{figure*}[ht]
\resizebox{\linewidth}{!}{    
\begin{tikzpicture}[
   node distance = 5mm and 7mm,
      start chain = going right,
 round/.style = {draw,
    rounded corners=0.5cm,
    minimum width=2.5cm,
    minimum height=1cm,
    text width=30mm, align=center, font=\linespread{0.8}\selectfont},
    disc/.style = {shape=cylinder, draw, shape aspect=0.3,
                shape border rotate=90,
                text width=16mm, align=center, font=\linespread{0.8}\selectfont},
  rect/.style = {draw,
    rounded rectangle,
    rounded rectangle arc length=45,
    minimum width=2.5cm,
    minimum height=1cm,text width=30mm, align=center, font=\linespread{0.8}\selectfont},
  alg/.style = {draw, align=center, font=\linespread{0.8}\selectfont},     arrow/.style = {thick, -Stealth},
  decision/.style = {diamond, aspect=1.5, draw, fill=green!30,
                    minimum width=3cm, minimum height=1cm, align=center,
                    on chain, join=by arrow},
                    ]
\node (n0) [round] {Select and prepare structures for workflows};          
\node (n1) [round,right = 10mm of n0] {Start with defaults in HT Workflows};
\draw[arrow] (n0.east)  -- (n1.west);
\node (n2) [disc,right = 10mm of n1] {Initial Results Data};
\draw[arrow] (n1.east)  -- (n2.west);
\node (n3) [round,right = 10mm of n2] {Perform Model selection and HPO};
\draw[arrow] (n2.east)  -- (n3.west);
\node (n4) [decision,right = 10mm of n3] {Model prediction \\more accurate than \\initial guess on \\test set?};
\draw[arrow] (n3.east)  -- (n4.west);
\node (n5) [round,right = 10mm of n4] {Proceed with integrated ML};
\draw[arrow] (n4.north) |- ++ (0,0.00003) -|  node[pos=.25,above]{No}  (n1.north);
\draw[arrow] (n4.east)  -- node[pos=.35,above]{Yes} (n5.west);
    \end{tikzpicture}\unskip}
\caption{Depiction of the initial data collection process and the decision path on when to proceed with DFT IntML. Cylinders denote a stage where data is processed. The green diamond denotes the decision in the workflow when to start with DFT IntML.}\label{Start}
\end{figure*}

Figure~\ref{Start} illustrates the workflow for obtaining initial data and the decision-making on when to continue with DFT IntML instead of other not data-driven parameter estimation methods.

\subsection{Data Requirements}
ML models are trained on data sets covering a certain subspace of the phase space. It is important to ensure that the phase space is properly sampled by the training data we want to use. Otherwise, the model will predict values outside the range it has been trained on. One can ensure this in DFT IntML by randomly sampling the calculations' phase space. However, this can be omitted if a batch is always used for DFT computation one after the other. However, the sampling is crucial if DFT IntML is performed iterative with very small batch sizes. 
\vspace{1cm}

\section{Results \& Discussion}

\subsection{Relaxed Structures without ML use}\label{Prob1}
Using the initial guesses described in section~\ref{MatProjGuess} and~\ref{magini} for the initial ILDs and the magnetic layers moments, we achieved convergence for 4316 different film systems. This corresponds to a convergence rate of $64.8\ \%$. Typically, failure rates of $10$ to $15 \ \%$ are considered acceptable in a high-throughput setting~\cite{Brder:891865}.
\label{Prob2}
There are multiple reasons that caused the relaxations to fail. As two loops are contained in the relaxation workflow (as described in Fig.~\ref{RelaxWF}), both can cause a failure. Hence, errors originate from both parts of the relaxation workflow, which includes:
\begin{itemize}
    \item The SCF calculations.
    \item The structural adjustments according to the determined forces.
\end{itemize}
Errors, originating from the relaxation process, typically show one of the following patterns:
\begin{itemize}
    \item MT spheres crash into each other as they relax too close for the given setup.
    \item MT spheres drift into the vacuum outside the film as the outer layer relaxes outwards too far away for the given setup.
\end{itemize}
Errors originating from the SCF calculations commonly result from the complex energy landscape and a difficult initial setup, as no minimum energy is found even with many iterations and no converged charge density is reached.
\subsection{Application of DFT Integrated Machine-Learning}
The standard approach to tackle a problem like the one mentioned in section~\ref{Prob1} would be a “trial and error” based change of input parameters for the failed relaxations. However, as it is hard to tailor a good “try” on this scale of failed relaxations for every system, a more systematic approach would be favorable as this would save even more computing time by avoiding unsystematic “tries”. 

Hence, we separated the atomic magnetic moments' data, which has been acquired using the relaxations of the ferromagnetic uniformly initialized film systems, into an 80/20 train/test split and performed a model selection, during which we found that XGBoost~\cite{XGBoost} regression is describing the data best of the tested models and then later on optimized the hyperparameter set involving a 4-fold cross-validation based approach on the training set. Our evaluation metric used was the MAE because we are not particularly concerned about a few large outliers but rather a small absolute error on most systems.

For this model, we were able to use very minimal input parameters. We used the atomic numbers of the magnetic layers and the substrate, which adds up to 4 integer features.\footnote{We chose that an unoccupied site corresponds to an atomic number of 0 in this representation.}
\subsubsection{First DFT IntML Batch}
In the first DFT IntML batch, we used the data obtained without the use of ML but with the structural and magnetic parameter guesses as described in the sections~\ref{MatProjGuess} and~\ref{magini}. With this data, we trained an XGBoost regressor model to predict the magnetic moments of the magnetic layer atom sites, given the atomic numbers of these atoms and the atomic number of the substrate material. 
The magnetic moments of the systems that did not converge using the initial structural and magnetic guess have been predicted using the model after it has been retrained on the entire available data set.
 Using this approach, one naively would expect a decreased MAE on the predictions compared to the relaxation outcomes. However, this is not the case, as shown in Fig.~\ref{RealWorld}. The reason for this can be found in Fig.~\ref{EvolOverData.png} from which it is apparent that the number of data points acquired using the initial ILD and magnetic moment guesses is already significantly higher than the mentioned break-even point and the incremental improvement per additional data point has already slowed down significantly. 
The predicted moments were then used as an improved initial starting point for the film system's SCF calculations.  
\label{fBa}
\subsubsection{Following Batches}
Improving the initial magnetic starting point only leads to the convergence of 570 additional systems relaxations. However, changing the initial magnetic moment only addresses parts of the previously described problems, which is why in the following batches, the procedure described in section~\ref{fBa} was extended to include a prediction for the ILDs of the magnetic layers. 
Optimizing the initial structure in addition to the initial magnetic moment not only provides us with a better starting point for the SCF calculation but also with a better starting point for the relaxation process closer to the relaxed structure. Hence, improving both the magnetic moment and the structural setting at the same time should lead to a significant boost in overall convergence rates. 

\begin{figure}[ht]
\includegraphics[width=\linewidth]{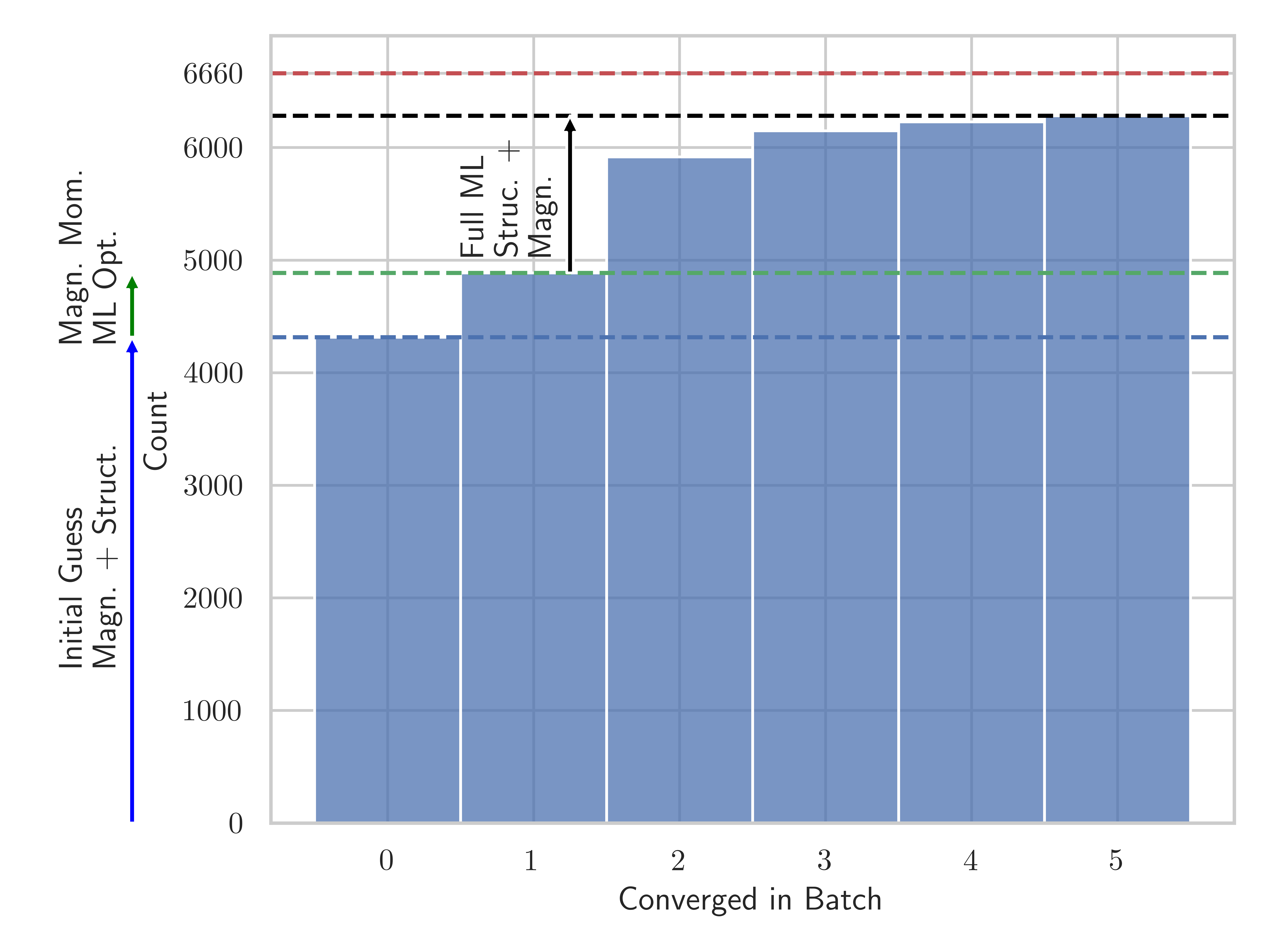}
\caption{Cumulative development of the number of converged relaxation for the different batches, including labeled arrows which indicate which degree of DFT IntML had which impact.}\label{Cumul}
\end{figure}

\begin{figure*}[btp]\centering
\includegraphics[width=0.7\linewidth]{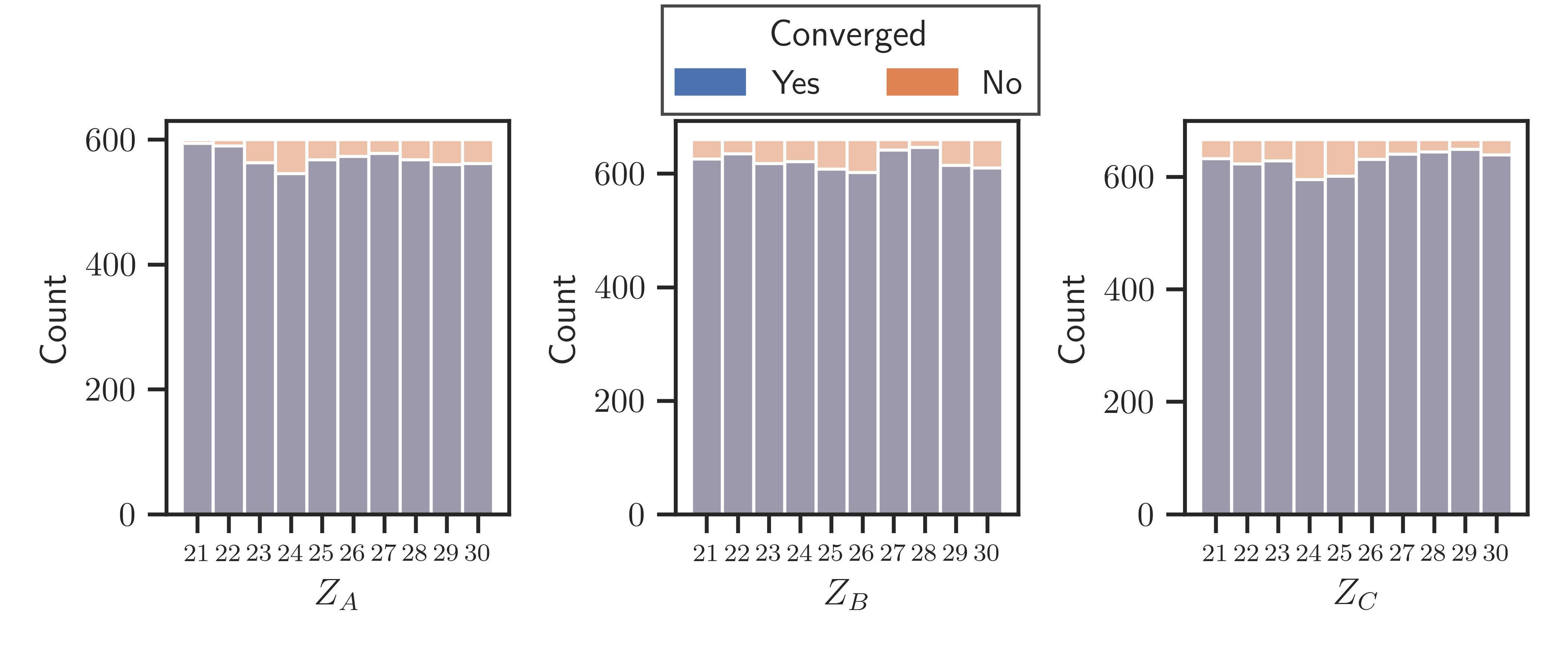}\\\includegraphics[width=0.45\linewidth]{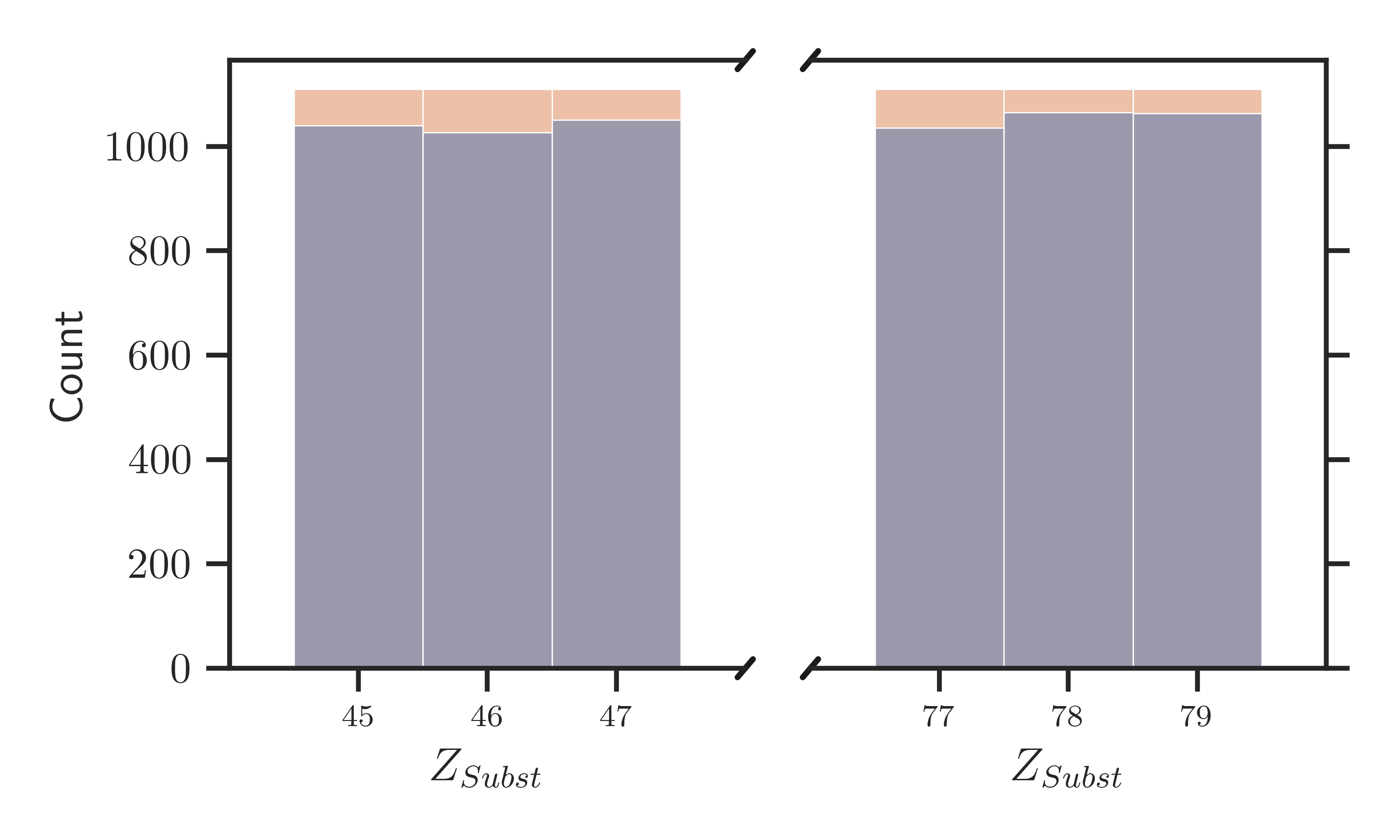}
\caption{Distribution of converged site occupations in relation to the different elements occupying these sites.}\label{distConv}
\end{figure*}

The development of the number of converged film systems over the batches is depicted in Fig.~\ref{Cumul}.
The DFT IntML approach boosted the number of converged systems to 6282 converged films, which accounts for $94.3\ \%$ of all the systems—leaving us with an error margin slightly above $5\ \%$, which is an excellent result for a magnetic high-throughput calculation of film systems. It also demonstrates the capabilities of the FLAPW method in a high-throughput setting despite having the peculiarity that MT spheres influence the relaxation process.   
Fig.~\ref{distConv} shows the distribution of atomic numbers and the converged fraction of the films containing the corresponding elements at the different layer sites.  

Fig.~\ref{distConv} shows that chromium and manganese in the A layer, iron and cobalt in the B layer and manganese and iron in the C layer are the atoms which seem to be hard to converge at the respective sites. All the substrates appear as equally challenging to converge successfully.

\subsection{Reduced computational time}
\begin{figure}[ht]
\includegraphics[width=0.9\linewidth]{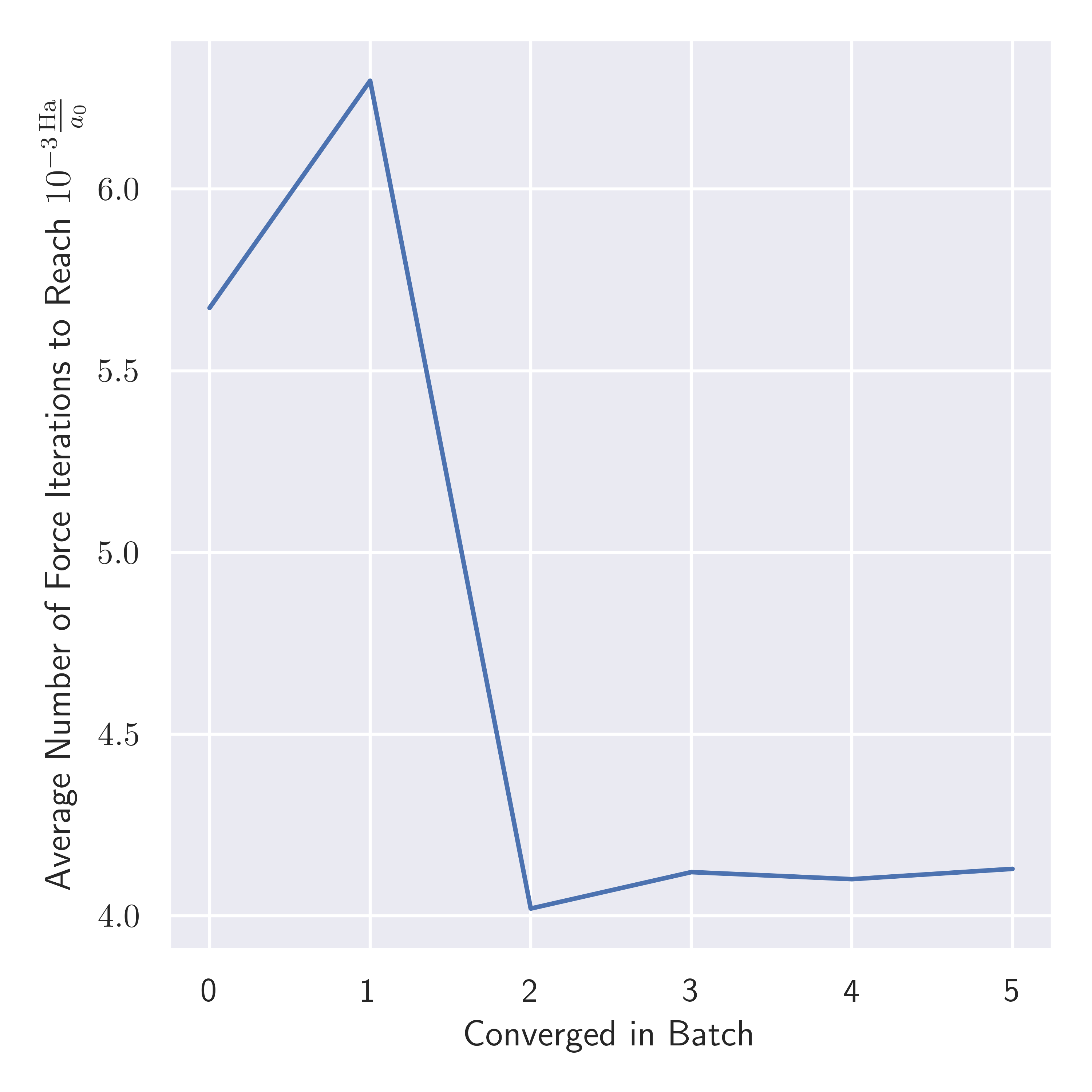}
\caption{Average required number of force iterations to reach a force threshold of $10^{-3}\frac{\mathrm{Ha}}{a_0}$ for each batch.}\label{improvB}
\end{figure}

\begin{figure}[ht]
\includegraphics[width=0.9\linewidth]{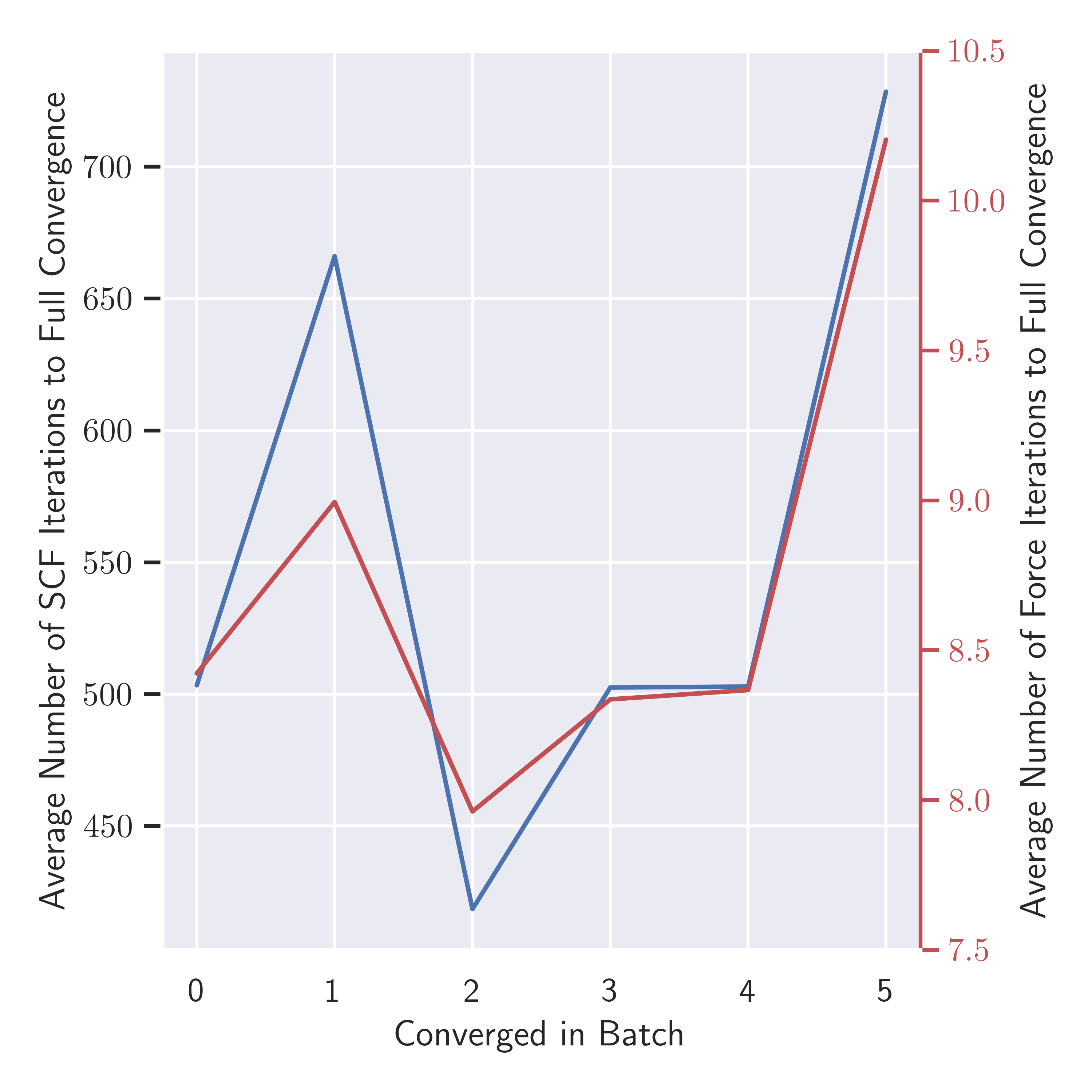}
\caption{Average required number of force iterations to reach a force threshold of $5\cdot 10^{-5}\frac{\mathrm{Ha}}{a_0}$  for each batch (red) and also the required number of accumulated SCF iterations to reach this convergence goal for each batch (blue).}\label{improvA}
\end{figure}

Besides the fact that the overall convergence rate is increased using DFT IntML, optimized starting points for the relaxation process \textit{i.e.}a predicted structure is likely to be closer to the actual relaxed structure. It will also require fewer relaxation steps to reach the relaxed structure. Fig.~\ref{improvB} shows the average number of relaxation steps which is needed to reach a maximum absolute force threshold of $10^{-3}\frac{\mathrm{Ha}}{a_0}$ for every batch. 
Fig.~\ref{improvB} indicates that setting an improved starting point for the SCF calculation using an optimized magnetic moment only in batch 1 first leads to an increased amount of required force iterations. This can be explained as only improving the initial magnetic moment may cause SCF calculations to converge, which did not happen beforehand and were more challenging to converge. However, this does not provide a better starting point for the relaxation. A significant drop can be observed after including the ML-optimized structural quantities in the DFT IntML workflow compared to the ML agnostic initial data. The relative drop from the initial data to the complete DFT IntML-based input optimization is about $27 \%$ of the initial required number of force iterations. Also, a reduction of, on average, up to $17\ \%$ of the total number of required SCF iterations to relax a film system could be achieved by ML optimizing both the structure and the magnetic layer moments, as shown in Fig. \ref{improvA}. While the reduced number of relaxation steps directly translates to fewer calculations necessary to relax a system, the number of SCF iterations required is directly proportional to the computing time used during the relaxation procedure. Hence, the DFT IntML method has been demonstrated to be capable of reducing the average computing time, the average number of submitted jobs, and the caused up- and download data traffic on a machine while at the same time improving the workflow's success rates due to the optimized input parameters.

However, enforcing our strict convergence criterion of $5 \cdot 10^{-5}\frac{\mathrm{Ha}}{a_0}$ for the force threshold, a similar plot also including the mean number of total SCF iterations is shown in Fig.~\ref{improvA}. In this plot, a very similar trend can be observed. However, one can also observe that reaching relaxation convergence is becoming a more challenging task with each additional batch. This can be explained as in batch five, only about $10\ \%$ of the converged systems could be described as ferromagnetic states. The other $90\ \%$ were classified as 
ferrimagnetic, antiferromagnetic, and non-magnetic states, which could be considered further located from the state we initially assumed at the beginning of the ML agnostic batch.

\subsubsection{Comparison of DFT IntML and guess errors}
\begin{figure}[ht]
\includegraphics[width=\linewidth]{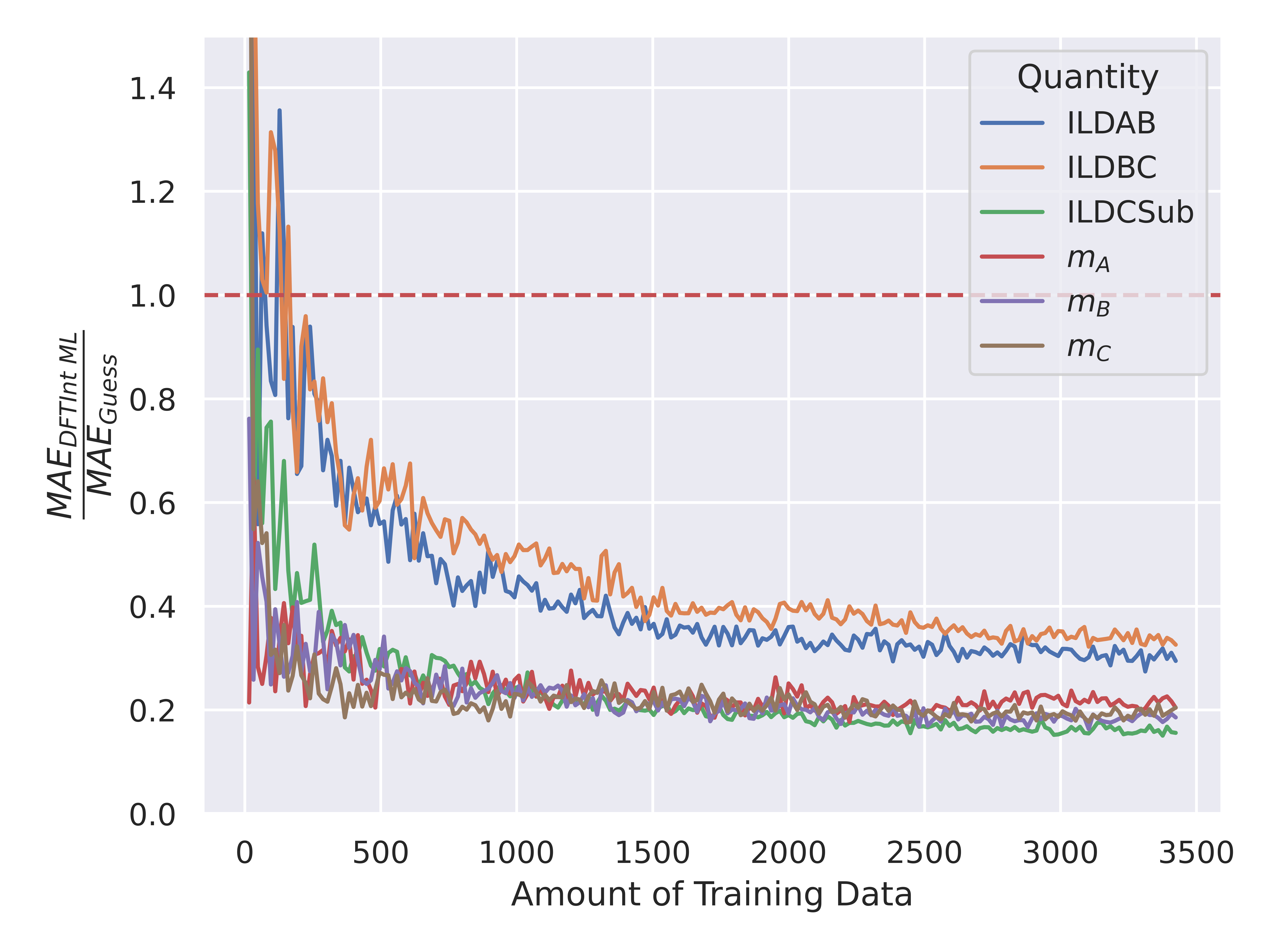}
\caption{Comparison of the errors from the initial guess and the posterior DFT IntML prediction as a function of the data amount accumulated. This posterior analysis has been compiled using the data from the initial batch. }\label{EvolOverData.png}
\end{figure}

Previously, we already mentioned that the best starting point for a DFT IntML approach is as soon as the ML prediction of the target quantity outperforms the data agnostic guessing method, which is used to obtain converged ab-into results initially. We can calculate the mean error of our guessing method (See table~\ref{GuessError} for the MAE values over the entire initial batch) and compare our DFT IntML approach as a function of the accumulated training data. This comparison is shown in Fig.~\ref{EvolOverData.png}.
\vspace{1cm}\\
\begin{adjustbox}{max width=\columnwidth}
\centering
          \begin{threeparttable}[b]
          
            \caption{Mean absolute error of the initial guessing method over the initially computed structures.}
          \label{GuessError}\centering
            \begin{tabular}{llll}
              \toprule
              Quantity & MAE & Unit\\
              \midrule
               ILDAB  & 0.092&\AA \\
               ILDBC  & 0.093&\AA\\
               ILDCSub  & 0.073&\AA\\
               Mag. Mom. A & 0.997 &$\mu_B$\\
               Mag. Mom. B  & 0.979&$\mu_B$\\
               Mag. Mom. C  & 1.018&$\mu_B$\\
              \bottomrule
            \end{tabular}
          \end{threeparttable}
        \end{adjustbox}
\vspace{1cm}\\
As the error from the initial guessing method is independent of the number of \textit{ab initio} calculations which we performed, the averages shown in table~\ref{GuessError} are enough to assist with the interpretation of Fig.~\ref{EvolOverData.png}.
The red line in Fig.~\ref{EvolOverData.png} indicates where the errors are equally large – meaning when the curves lower below the red line, the DFT IntML approach outperforms the initial guesses. For this posterior evaluation, the data has been sampled randomly. The train/test split is 80/20. The prediction error has been evaluated on the test set only. 

From Fig.~\ref{EvolOverData.png} one can see the break-even point between the guessing method for both the initial magnetic moment and ILDs and the DFT IntML approach after not even 300 data points. One can also see that the error in the DFT IntML method continues to improve even in the regions where most of the data has already been accumulated, even though the incrementally increased accuracy per additional data point decreases as more data is gathered for model training. The MAE from DFT IntML predicting the ILDs and magnetic moments compared to the DFT results is between $66\ \%$ and $80 \ \%$ smaller than the guessing error of the initial parameter guessing methods in this posterior model analysis. 
This indicates that an early start of the DFT IntML scheme benefits the convergence rate due to improved starting parameters provided to the \textit{ab initio} calculations and workflows. However, while an early start of the DFT IntML scheme can benefit the convergence rate, the MAE development of the model shows significant improvements for the training data amounts close to the break-even point (red line in Fig. \ref{EvolOverData.png}), which implies that when using DFT IntML from the break-break even point to the initially used guessing methods, small batch sizes, and frequent retraining, taking into account the additionally acquired data, can be beneficial.

\begin{figure*}[ht]%
\includegraphics[width=0.5\linewidth]{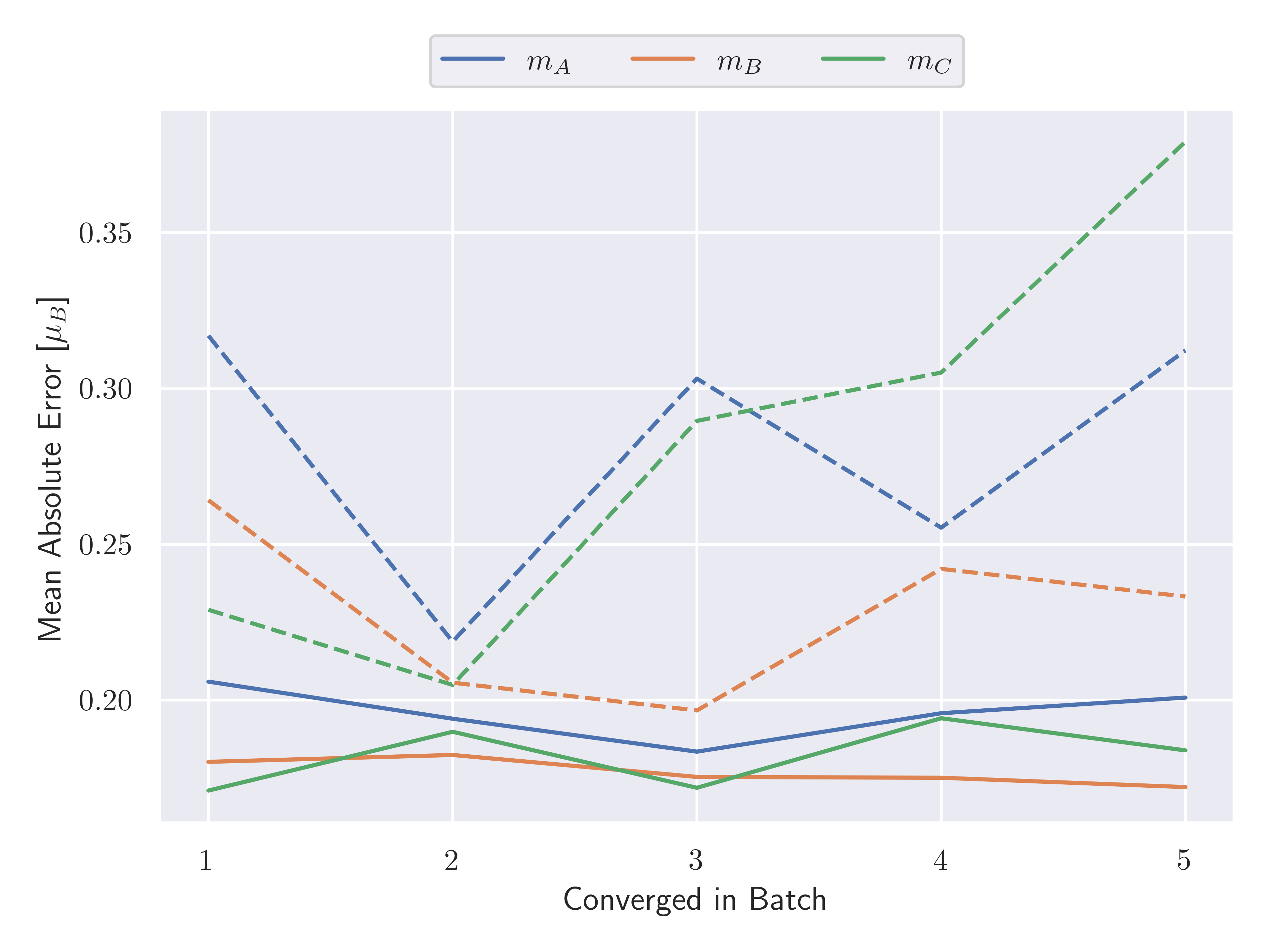}\includegraphics[width=0.5\linewidth]{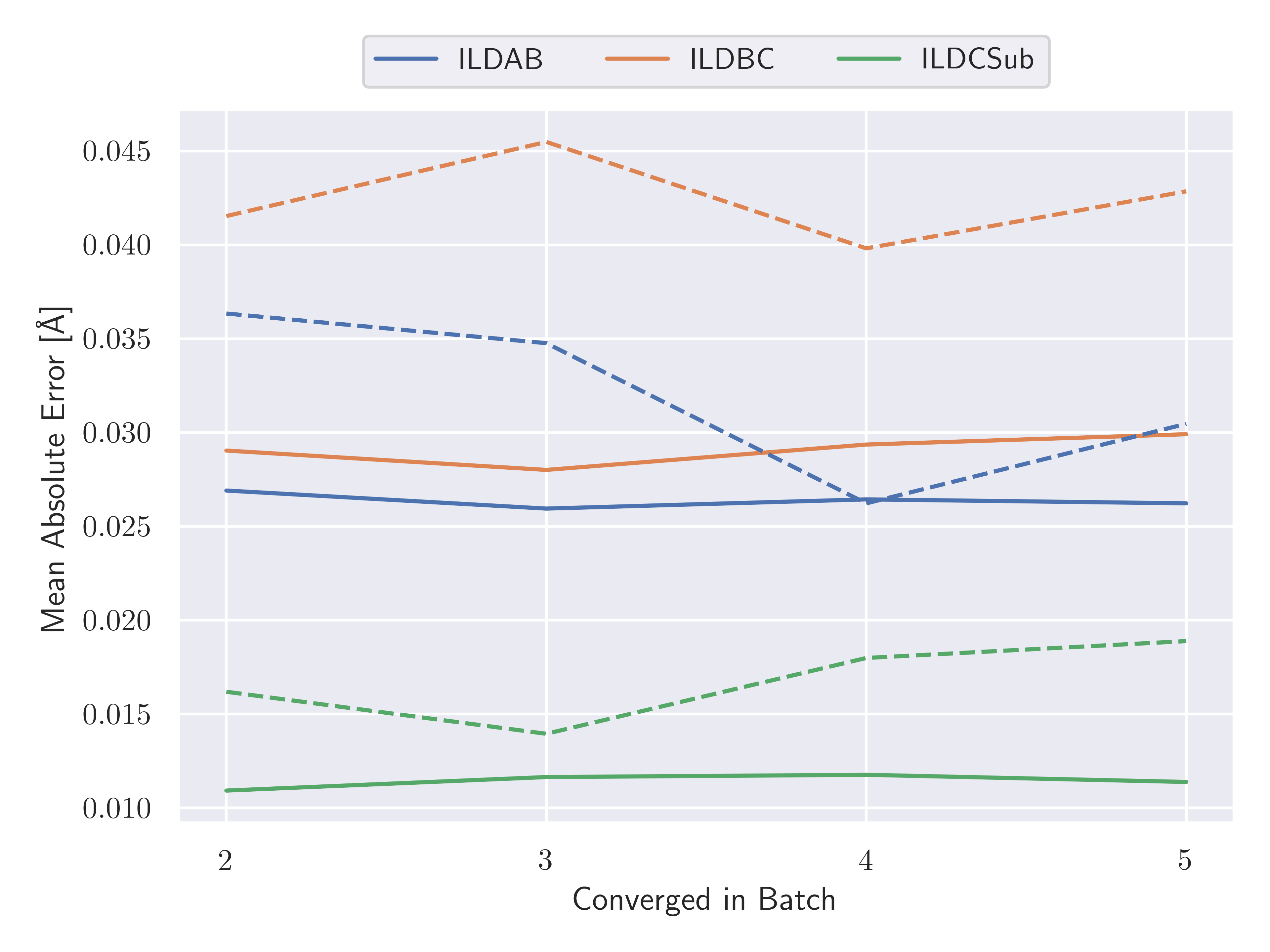}
\caption{Depiction of the test set MAE development as a function of the corresponding batch in comparison to the MAE of the actual predictions, based on the training data from the previous batches, which converged in the corresponding batch. The dashed lines represent the errors of the actual predictions, while the continuous lines represent the posterior determined prediction error.}\label{RealWorld}
\end{figure*}

However, due to the randomization we applied to the whole data set – and hence mixing all batches together – the previous discussion follows a few assumptions when performed as a batch learning process: 
\begin{itemize}
    \item Different batches are comparably challenging to predict. (Similar modeling  complexity)\footnote{This also includes that no data islands exist. \textit{e.g.} some films seem to follow fundamentally different underlying mechanics than most of the data set.}
    \item Each following batch samples the remaining phase space equally well as the previous one. (Sampling quality)
    \item Calculation parameters do not change from batch to batch (or even inside batches) except from the predicted input quantities. (Parameter independence)
\end{itemize}
While the test MAE values for an ML model can be considered a rough estimation of the potential prediction error, using the MAE as such requires that the not converged structures have a similar prediction complexity as the randomly chosen test set, which itself implies that the test set samples the phase space of all structures examined in the high-throughput study appropriately. 
However, posteriorly, we can evaluate if this was the case.
 Since we stored our prediction of each batch for every quantity, we can examine if this aligns with the error development of our 4 DFT IntML batches. The corresponding MAE scores are shown in Fig.~\ref{RealWorld}.
From Fig.~\ref{RealWorld}, it is clear that the naive expectation that the real prediction error would drop below the test set error as we use the whole data set for training is not correct. This has a few reasons: 
\begin{itemize}
    \item The modeling complexity differs for different batches. Later batches are expected to contain a larger fraction of compounds that are more difficult to model. This can already be seen in Fig.~\ref{improvA} as the last batch, on average, requires a lot more total iterations than the previous ones. 
    \item We can, of course, impact the sampling of the phase space in a DFT IntML workflow with very small batches, and we highly recommend doing so. However, even though we sample the remaining phase space for each batch, we can not control if the compounds converge in an order that samples the phase space homogeneously. 
    \item Changing the ILDs also changes the structural setup performed in the \texttt{\uppercase{Fleur}} code. This includes the total film thickness and the muffin-tin radii. 
\end{itemize}
Hence, in a real-world application, a guess improvement of $66 \ \%$ to $80 \ \%$ as previously discussed is unlikely. However, taking into account the values from table~\ref{GuessError}, which represent the guessing method MAE values acquired by comparing the initial guesses to the converged results, we are left with a minimum prediction error reduction of around $ 50 \ \%$ to $60\ \%$ for both the magnetic moments and the ILDs compared to the guessing error. However, it is important to keep in mind that this improvement was enough to reduce the required number of relaxation steps and SCF iterations significantly and that this improvement is possible for the estimation of optimized input ILDs even though a considerable effort has been made to find suitable starting ILDs using the average bond length estimation method provided within AiiDA-FLEUR~\cite{aiidaFleur}. 
\section{Summary \& Outlook}

From the previous observations, it is clear that the traditional trial and error approach to input optimization, typically used to improve the success rate of high-throughput \textit{ab initio} studies, can be replaced with a systematic ML-based approach. This approach is not limited to magnetic moments or ILDs but applies to any quantity, which is both the input and output quantity of a DFT calculation. Examples of other quantities that could be optimized this way would be \textit{e.g.} bond lengths in general, non-collinear magnetic moment orientation angles, and the charge density itself. Additional features beyond the constituent's atomic numbers should be considered for other applications. 
\\ Considering the benefits that were measurable in the presented application with the use of DFT IntML optimized structural and magnetic inputs values, which include a reduction of on average $17 \ \%$ of SCF iterations, a $29 \ \%$ reduction of needed average relaxation steps, significantly ($50\ \%$ to $60\ \%$ compared to the relaxation results) improved inputs to acquire a relaxed film, and an increased overall thin film structure convergence rate by nearly $30\ \%$ up to $94.3\ \%$, we see potential in the presented methodology to assist with common issues arising during fist-principles high-throughput studies beyond the presented application. 
\\ Adapting the methodology of batch learning and hence integrating ML into high-throughput applications and submissions scripts represents an example of lightweight and easily automatable ML methods that can assist within existing and established computational methods - such as DFT - to harness the availability of already computed data to benefit the high-throughput study itself and hence boost scientific discoveries beyond the existing data. 

\ack
This work was performed as part of the Helmholtz School for Data Science in Life, Earth and Energy (HDS-LEE) and received funding from the Helmholtz Association of German Research Centres.

 The authors gratefully acknowledge the computing time granted by the JSC (Project: fleur4thc) on the supercomputer JURECA-DC at Forschungszentrum J\"ulich and also the computing resources granted by RWTH Aachen University (Project: jara0234) on CLAIX.
 
Since parts of the data processing have been performed and the displayed visualizations have been created using dedicated open-source packages and tools, we acknowledge them here~\cite{numpy,mendeleev2014,Tikz,MPL,MPLV,Seaborn,XGBoost,scikit-learn}.
We thank Vasily Tseplyaev for his continuous support in technical questions using the AiiDA-FLEUR framework and workflows during - and even after - his PhD project. 
This project used dedicated software tools, which are materials science and/or \textit{ab initio} specific. We acknowledge them here~\cite{Masci,MaterialsCloud,MatProj,aiidaFleur,AiidaA,AiidaB}. An essential mention in this context is Henning Janssen, who assisted us with the use of Masci-tools in the early stages of this study.

\section*{References}
\bibliographystyle{iopart-num}
\bibliography{Paper.bib}

\end{document}